\documentclass[journal=langd5, manuscript=article]{achemso}
\usepackage[english]{babel}
\usepackage[usenames]{color}
\usepackage{ulem}

\usepackage{graphicx}
\usepackage{bm}
\usepackage[version=3]{mhchem} 
\usepackage[sort&compress]{natbib}

\title{Solvent contribution to the stability of a physical gel characterized by quasi-elastic neutron scattering}

\author{Sylvie Spagnoli}
\affiliation[UJF-LIPhy]{Univ. Grenoble Alpes, LIPhy, F-38000 Grenoble, France}
\alsoaffiliation[CNRS-LIPhy]{CNRS, LIPhy, F-38000 Grenoble, France}
\author{Isabelle Morfin}
\affiliation[UJF-LIPhy]{Univ. Grenoble Alpes, LIPhy, F-38000 Grenoble, France}
\alsoaffiliation[CNRS-LIPhy]{CNRS, LIPhy, F-38000 Grenoble, France}
\author{Miguel A. Gonzalez}
\affiliation[ILL]{Institut Laue Langevin, BP 87, 38042 Grenoble Cx 9, France}
\author{Pierre \c Car\c cabal}
\affiliation[ISMO]{Institut des Sciences Mol\'eculaires d'Orsay, CNRS and Universit\'e Paris Sud, Orsay, F-91405, France}
\author{Marie Plazanet}
\email{marie.plazanet@unipg.it}
\affiliation[UJF-LIPhy]{Univ. Grenoble Alpes, LIPhy, F-38000 Grenoble, France}
\alsoaffiliation[CNRS-LIPhy]{CNRS, LIPhy, F-38000 Grenoble, France}
\alsoaffiliation[Universit\`a di Perugia]{Dipartimento di Fisica, Universit\`a degli Studi di Perugia, I-06123 Perugia, Italy}

\begin{document}
\date{\today}
\begin{abstract}
The dynamics of a physical gel, namely the Low Molecular Mass Organic Gelator {\textit Methyl-4,6-O-benzylidene-$\alpha$ -D-mannopyranoside ($\alpha$-manno)} in water and toluene are probed by neutron scattering. Using high gelator concentrations, we were able to determine, on a timescale from a few ps to 1 ns, the number of solvent molecules that are immobilised by the rigid network formed by the gelators. We found that only few toluene molecules per gelator participate to the network which is formed by hydrogen bonding between the gelators' sugar moieties. In water, however, the interactions leading to the gel formations are weaker, involving dipolar, hydrophobic or $\pi-\pi$ interactions and hydrogen bonds are formed between the gelators and the surrounding water. Therefore, around 10 to 14 water molecules per gelator are immobilised by the presence of the network. 
This study shows that neutron scattering can give valuable information about the behaviour of solvent confined in a molecular gel.
\end{abstract}

\section{Introduction}
Physicals gels formed by low molecular mass organic gelators  (LMMOG) represent very interesting materials, both for fundamental physics as well as applied science. They are composed of a rigid network formed by the gelators, in which is trapped a large quantity of solvent. Gathering properties of both liquids and solids, they find applications in diverse domains, such as temperature sensors or drug delivery \cite{Sangeetha2005,Vintiloiu2008,Valery2003,Pozzo1998}. They present a high potential for functionalised nanomaterials, thanks to the versatility of functionalized molecules that can be used as gelators\cite{Terech1997,DeJong2004,Gronwald2002,vanEsch2000}, and the supramolecular organisation that can be tuned by changing the solvent. 

The same gelator can indeed present various structural organisations, like ribbons, vesicles or even nanotubes\cite{Boettcher2001, Cui2008}, that may arrange in periodic uni- or bidimensional structure. The particular arrangement adopted is defined by a subtle interplay between different forces like hydrogen bonds, hydrophobic forces, $\pi-\pi$ interactions or dipolar forces, so it will depends on the solvent and will be often restricted to a particular temperature range. The gelation of new molecules, even after intensives research based  on steric effects and crystalline structure remains poorly predictable\cite{Luboradzki2000, vanEsch2000, Weiss2009}. Moreover, applications suffer several limitations, due to the metastability of the gels that may collapse after some delay, spanning from the hour to the month or longer. Although having certain organisation, it is known that these systems are rather disordered and can show several relaxation modes related both to the gelators and the solvent dynamics. 

If structural information on LMMOG  obtained by Transmission Electron Microscopy (TEM) or small angle scattering is often available, this is not the case concerning the dynamics. Their stability can be evaluated from the gelation temperature and melting enthalpy \cite{Yoza1999, Bielejewski2011, Bielejewski2009} and compared to solvent parameters characterizing its polarisability or hydrogen bond ability \cite{Hirst2004, Edwards2011, Bielejewski2011}, leading to a phenomenological description of the gelator-solvent interactions. Infra-Red spectroscopy can be used in order to investigate the strength of hydrogen bonding \cite{Yoza1999}, related to the stability of the network.
The dynamics of both gelators and solvent molecules are important contributions to the gel stability, as a source of entropy. The solvent dynamics should be influenced by the presence of the gel, specially if some solvent molecules participate into the rigid network. In that case, several solvent populations may be distinguished. The gelators mobility is also greatly affected by the gel formation. Mainly NMR-based techniques have been used to quantify the mobility of the gelators and the solvent in both the sol and gel phases \cite{Shapiro2011, Tritt-Goc2008,Yoza1999,Tritt-Goc2011, Tritt-Goc2012}, on timescales slower than tens of nanoseconds. 
The solubility of the gelators in the solvent, which is small but finite, leads to a dynamical exchange of gelators between the rigid network and the liquid part. The network is therefore a dynamical assembly \cite {Duncan2000}. The life time of a gelator in the rigid network is expected anyway to be longer than the nanosecond \cite{Duncan2000, Shapiro2011}, so gelators appear immobile on a shorter timescale.

Given the typical gelators concentration and sample slight turbidity, the characteristic size of the pores is estimated to be larger than a few hundred nanometres. A large part of the solvent is therefore expected to present a bulk-like behaviour. However, the interaction between the gelators and the solvent molecules leads to an organisation of the solvent close to the rigid parts, as well as a slowing down of its dynamics \cite{Vogel2010, Busselez2009, Liu1991}. Such behaviour is commonly observed for liquids confined in matrices. Water confined in porous silicas is possibly the best characterized example of this kind of system and it has been shown that even in pores of $\sim$ 2 nm of diameters, only 2 water layers are influenced by the matrix, and the third layer presents  already a bulk-like behaviour\cite{Taschin2013}. A similar behaviour is found for confined solvents in hard or soft confinement matrices\cite{Busselez2009}. 
The dynamics of solvent is expected to exhibit a continuous transition \cite{Corkhill1987} between the most hindered (interacting) solvent molecules and the bulk-like ones. In practice, this behaviour is experimentally characterized by the distinction of several populations, varying between 2 and 4 depending on the experimental techniques \cite{Pissis2013}. In molecular gels, the mobility of the solvent has been investigated by NMR \cite{Duncan2000,Shapiro2011} on timescales slower than 20 ns. Quasi-elastic Neutron Scattering operate on a faster timescale and is therefore particularly well suited for characterising the faster solvent dynamics, but it has barely been used to investigate the dynamics in gels, as the solvent contribution usually overwhelms the whole signal \cite{Seydel2008}.


In this context, we undertook the study of the microscopic dynamics in a model physical molecular gel, with the goal of characterising the molecular interactions and understanding the formation and stability of the gel phase. As model system, we chose to investigate a gelator that belongs to a sugar-based gelator family first reported by the group of Shinkai and collaborators, methyl-4,6-O-benzylidene-$\alpha$-D-{\bf mono}pyranoside \cite{Gronwald2001}. It is one of the simplest organic molecular gelators family described in the literature, made of a saccharide carrying an aromatic benzylidene group as schematized in the Scheme \ref{manno}. The methyl-4,6-O-benzylidene-$\alpha$-D-glucopyranoside ($\alpha$-gluco) and methyl-4,6-O-benzylidene-$\alpha$-D-mannopyranoside ($\alpha$-manno) derivatives present the best gelation properties. They only differ by the orientation of one of the OH groups (see Scheme \ref{manno}), but the $\alpha$-manno gelifies protic solvent (water) as well as apolar organic solvents (toluene), while the $\alpha$-gluco does not gelify water. $\alpha$-manno is therefore a small molecule carrying not more than the necessary functions for an amphiphile gelator. Its compact shape enables to use simple models for its dynamical relaxations (rotational and translational diffusions) and possible interactions with the solvent. 

Like most gelators,  $\alpha$-manno adopts different supramolecular conformations depending on the solvent, forming more or less complex assemblies. For example, a remarkable structure is the one formed in p-xylene, in which fibres of $\alpha$-manno are organized in an hexagonal structure with characteristic distances of 60 \AA , involving solvent molecules in the rigid network \cite {Sakurai2003}. As a general trend, the gelators assemble in water through the stacking of the benzylidene parts and form hydrogen bonds with the solvent while in apolar or weakly polar solvents, the gel formation is driven by intermolecular hydrogen bonds between the gelators and the aromatic groups are solvated by the liquid part \cite{Cui2008, Gronwald2001a}. 

With the goal of extending the investigated timescale of the dynamics in molecular gels, we performed quasi elastic neutron scattering on the model system $\alpha$-manno in water and toluene. We aim at characterizing how the solvent dynamics are affected by the gelification and how it compares in different kind of solvents.  Combining Time Of Flight and Backscattering data, we are able to span the dynamics over four orders of magnitude, from tenth of picoseconds up to the nanosecond in order to characterize the various solvent populations. In all our measurements, given the characteristic timescales previously discussed, the gelators are therefore considered to be immobile for translation.
After the overview of the Material and Methods, we will present the experimental results and analysis, and discuss them in regard of results from the literature obtained by NMR techniques.


\begin{scheme}
\includegraphics[width=8cm]{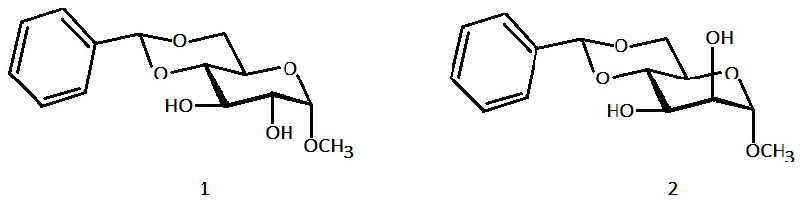}
\caption{methyl-4,6-O-benzylidene-$\alpha$-D-glucopyranoside (1) and methyl-4,6-O-benzylidene-$\alpha$-D-mannopyranoside (2)} \label{manno}.
\end{scheme}

\section{Materials and Methods}
\label{sec:Material}
{\bf Sample preparation.} $\alpha $-manno ($C_{14}H_{18}O_6$) was purchased from Orgentis Chemicals and dried under vacuum. Organic solvents were bough from Sigma-Aldrich, distilled at $\sim 80^O C$ and dried over molecular sieve. $\alpha$-manno powder was mixed with water or solvent and heated to $ 80^O C$, and the gels were formed during cooling at room temperature. 
The Table \ref{tbl:samples} describes the different samples used on both instruments. In order to check the concentration dependence of the network constituents, we performed the measurements on IN5 (see next paragraph) on three different contributions of $\alpha$-manno toluene gels.
In all samples, the residual solubility of the gelators was always lower than $\sim ~10^{-1} g/L$, and therefore neglected as a contribution in the liquid part. We checked before and after all measurements that the gels did not collapse.

\begin{table}
   \begin{tabular}{|c | c | c | c | c | c |}
\hline
     instr. & solvent & [c] g/L & numb. solvent/gelator & I gelators & I solvent \\
	\hline     
     IN5 & toluene - T1 & 51.2   & 52.1 & 0.04 & 0.96\\
      & toluene - T2& 103.6   & 25.7  & 0.08 & 0.92\\
      & toluene - T3& 145.0   & 18.4  & 0.11 & 0.89\\
      & H$_2$O - 1  & 88.2   & 177.5 & 0.05 & 0.95\\
\hline
   IN16 & toluene - T4& 139.3  & 19.1 & 0.10 & 0.90\\
           & H$_2$O - 2  & 208.4 & 75.2 & 0.11 & 0.89\\
\hline
   \end{tabular}
\caption{$\alpha$-manno gels used in the QENS experiments and contribution of each component to the total neutron scattered intensity.}
\label{tbl:samples} 
\end{table}

{\bf Neutron Scattering.} Quasi elastic neutron scattering was performed at the Institut Laue-Langevin on the spectrometers IN5\cite{Ollivier2004} and IN16\cite{Frick2001} on fully protonated samples. On IN5, the quasi elastic signals were measured using a wavelength of $\lambda$=5 \AA, providing a resolution of 100 $\mu$eV and the Q range [0.2-2.2] \AA$^{-1}$ as well as a wavelength of $\lambda$=10 \AA, providing a resolution of 10 $\mu$eV and the Q range [0.1-1.1] \AA$^{-1}$. On IN16, the instrumental resolution was 1$\mu$eV and the Q range [0.2-1.9] \AA$^{-1}$. The quasi-elastic spectra were measured between -10 and 10 $\mu$eV, while the so-called fixed window scans enabled the monitoring of the (elastic) intensity integrated over the instrumental resolution in a temperature range of [50-350 K]. In these last experiments, the samples were first heated up to above the sol-gel transition, and then cooled down to low temperature, to avoid an eventual damage of the gel at low temperature. The sol-gel transition is indeed reversible, but the effect of the solvent crystallisation in the network may cause damages in the gel structure.

The Table \ref{tbl:samples} also reports the fraction of neutron scattered intensity of both constituent of each sample. The calculations are based on the incoherent cross section, widely dominating the whole signal: $\sigma_{\alpha-manno}$=1438.4, $\sigma_{toluene}$=639.3 and $\sigma_{water}$=159.8 barns, respectively for the $\alpha$-manno, the toluene and the water molecule. The data reduction and basic corrections were performed using the standard procedures in the LAMP program\cite{LAMP}. The IN5 data were rebinned at constant Q values, but for the analysis of the elastic intensities the constant 2$\Theta $ binning was used, enabling the use of a larger dynamical range. The sample thicknesses were of 0.2 mm, ensuring a transmission higher than 0.87 and enabling to neglect multiple scattering.

\section{Experimental results}

{\bf Fast solvent dynamics.} We first investigated the fast dynamics using Time of Flight measurements (IN5, ILL) with two different resolutions, giving access to the timescale 0.1-100 picoseconds. The bulk diffusion coefficients of water and toluene at 293 K are respectively around 2.0$\cdot 10^{-5}$ cm$^2$/s \cite{Qvist2011} and 2.7$\cdot 10^{-5}$ cm$^2$/s \cite{Klein2003}. The investigated dynamical window is therefore suitable to observe such translational dynamics of bulk solvents, as well as dynamics 10 to 15 times slower.

\begin{figure}
\includegraphics[width=10cm]{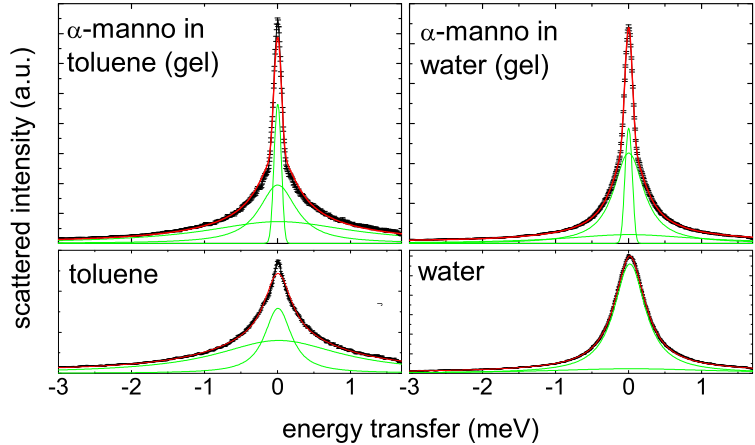}
\caption{Picosecond dynamics measured by QENS on the Time of Flight IN5 at 290 K, Q= 1.3\AA$^{-1}$ in the gel of $\alpha$-manno in toluene (T3) and water ($H_2O $-1), and pure toluene and water for comparison (bottom). The lines show the fitting to the experimental points using the equation (\ref{eq:eq_fit}) (red line) and the different elastic and quasielastic contributions (green lines)(see text).} \label{QENS_IN5}.
\end{figure}

Both gels and pure solvents were measured in the same conditions for comparison. Characteristics quasi-elastic neutron scattering spectra of the gels and corresponding pure solvents are shown in Figure \ref{QENS_IN5}. The main difference observed between the gels and pure solvents is an elastic contribution that emerges in the gel as a narrow peak on top of the broad quasi elastic contributions arising from the liquid fraction, evidencing the presence of the rigid network.

Both set of data, in water and toluene, were fitted with a phenomenological model described by the equation \ref{eq:eq_fit} and made of two lorentzian contributions and an elastic peak. This model is not the most sophisticated one that can be used to describe the dynamics of water and get a proper separation of the rotational and translational dynamics. However, it better suits our purpose, which is the differentiation of water dynamics in bulk and in gel and the comparison of the translational diffusion coefficient, enabling us to use the same model for all the samples investigated here. This theoretical form was further convoluted with the instrumental function, and no constraints were applied to the parameters.
\begin{equation}
S(Q,\omega)=\left( A_0(Q) \cdot \delta(\omega)+A_1(Q)L_1(Q,\omega)+A_2(Q)L_2(Q,\omega) \right) \otimes R(Q,\omega)
\label{eq:eq_fit}
\end{equation}
All the spectra exhibit a broad lorentzian having a width of several meV with a mild $Q$-dependence that is associated to rotational dynamics, and a narrower lorentzian (0.1-1 meV FWHM) showing the characteristic $Q$-dependence of a translational diffusion process.
The agreement with experimental data was satisfactory, and no additional contribution of intermediate relaxation time could improve the fit.

\begin{figure}
\includegraphics[width=8cm]{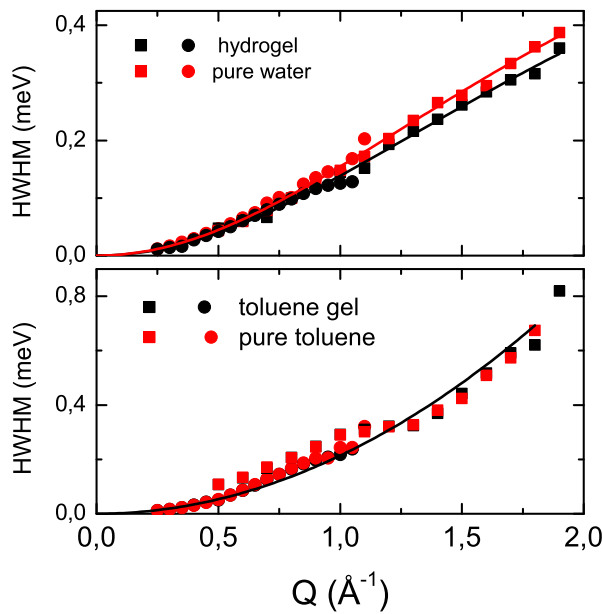}
\caption{FWHM of the translational component of the solvent signal in the pure solvents and associated gel measured on IN5 with a wavelength of 5 (squares) and 10 $\AA$ (circles). Top : $\alpha$ manno-$H_2O$ (black symbols) and pure $H_2O$ (red symbols), bottom :  $\alpha$ manno-Toluene (black symbols) and pure Toluene (red symbols). 
The fit of the jump diffusion in water samples and Fick diffusion for Toluene samples are represented as solid lines for comparison. The diffusion coefficients are discussed in the text.} \label{QENS_width}.
\end{figure}

{\bf Translational component}. The width of the translational contribution over the whole Q range is represented in Figure \ref{QENS_width} for the different samples and the two experimental configurations. In the three toluene samples, no difference can be observed between the gel and pure solvent, meaning that no slowing down of the solvent dynamics could be detected, whatever the gelator concentration is. More sophisticated fits using rotation-translation models, or taking extensively account of the solvent in the gel samples lead to the same result: no intermediate dynamics is observed between the bulk-like behaviour of the sample and the molecules having a relaxation time slower than 100 ps, and that are therefore detected as immobile, contributing to the elastic peak.

The toluene dynamics follows a Fickian behaviour, with a width of the quasi-elastic component increasing linearly with $Q^2$. We can however observe a deviation of this law at $Q \sim 1.2 $\AA$^{-1}$, corresponding to a distance of 5.3 \AA  which is close to the distance between the center of mass of the molecules. This plateau is typical of the De Gennes narrowing, that consists in a slowing down of the translational dynamics when the translational distance approach the equilibrium distance between the molecules. This effect is observed only in the coherent neutron scattering signal (8\% of the total scattering in the case of toluene), that increases at this momentum transfer according to the structure factor.
The toluene dynamics can however be well approximated by a simple Fickian diffusion, with a coefficient of 3.2$\cdot 10^{-5} cm^2.s^{-1}$. The fit of this coefficient is represented in Figure \ref{QENS_width} (bottom). This slight overestimation with respect to values reported in the literature is assigned to the use of separated components between rotation and translation, instead of performing a convolution between the two components as would be required by a more accurate treatment.

In water, the translational dynamics is modelled according to the jump diffusion behaviour \cite{Egelstaff1967}. The respective coefficients for pure water and hydrogels are D=2.8 $\pm$ 0.1$\cdot 10^{-5} cm^2.s^{-1}$ and D=2.5 $\pm$ 0.1$\cdot 10^{-5} cm^2.s^{-1}$, and the residence times $\tau$ = 0.75 $\pm$ 0.03 ps, and $\tau$ = 0.77 $\pm$ 0.04 ps. The translational dynamics of water is therefore sightly slower in the gel than in pure water. We note that the single-lorentzian character of the signal proves that the whole water is slowed down and not only a fraction of it that would be averaged with the other bulk-like fraction. The effect, as will be discussed in more details in the Discussion part, is however very small compared to what ca be observed on other systems.

{\bf Rotational component}. The large component is typical of rotation or localized motions, exhibiting a rather flat width and increasing amplitude when Q increases. In pure toluene, the component has a full width of 2.7 $\pm$ 0.3 meV, while the corresponding component in the gel, having the same characteristics at small Q, tends to increase for Q larger than 1.4 \AA$^{-1}$ up to $\sim$ 6 meV. The case of water gels is very similar: while the pure water exhibits a rotational component of constant $FWHM=1.5 \pm 0.2 $meV, the quasi elastic component width in the gel increases up to $\sim$ 6 meV. The increase of the component at high Q is interpreted as a quasielastic component arising from a localized motion of the gelators within the network, like  flip-flop motions of the sugar moieties or rotation of the aromatic groups.

{\bf Elastic component}. Although no influence of the network can be detected on the solvent mobility, quantitative information about the network composition can be extracted from the elastic peak, which arises from all the atoms in that system that do not diffuse or appear as immobile on the timescale given by the instrument resolution, i.e. t \textless 13 ps. 
In order to evaluate the elastic contribution, the fitting of the data measured in the gels was performed using the pure solvent as a (scaled) contribution. This way, only an additional elastic contribution (delta function) was necessary to fit the data. We should note here that, as stated in the discussion of the rotational component, a quasi-elastic signal arising from the network should be detected at high Q. However, it is not possible to reliably distinguish a new rotational component in the data. The small quasi-elastic signal of the gelators is therefore drown in the solvent rotational contribution and the error bars.  

The total scattering function may then be written as a sum of liquid and solid contribution. As previously shown by the analysis of the translational diffusion coefficients, the liquid parts exhibit diffusion coefficients which are highly similar to those of the bulk solvent, enabling us to identify the liquid contribution with the pure solvent one: $S(Q,\omega)=\alpha S_{solid}(Q,\omega)+(1- \alpha)S_{solvent}(Q,\omega)$. The solvent part is therefore purely quasi-elastic while the solid one is the sum of elastic and quasielastic terms due to localised motions of gelators or solvent molecules trapped in the rigid network. Moreover, the total scattered intensity is conserved such that $I_{el}+I_{qens}=1$. 
The quantity $\displaystyle R(Q)= \left(\frac{I_{el,solid}}{I_{el,solid}+I_{qens,solvent}}\right)$ recalls the Elastic Incoherent Structure Factor (EISF) usually defined in solids\cite{Bee1998}. 
 In the limit of $Q \rightarrow 0$, all the atoms belonging to the solid part, and only them, contribute to the elastic component, as their quasi-elastic contribution drops to 0. We therefore get an evaluation of the solid part from the ratio $\displaystyle R(0)= \left(\frac{I_{el,solid}}{I_{el,solid}+I_{qens,solvent}}\right)_{Q=0}=\left(\frac{\sigma_{\alpha manno}+n_{solvent, immo}\cdot\sigma_{solvent}}{\sigma_{\alpha manno}+n_{solvent, total}\cdot \sigma_{solvent}}\right)$, where $\sigma_{solvent}$ and $\sigma_{mannno}$ are the scattering cross sections and $n_{solvent, immo}$ and $n_{solvent}$ are the number of solvent molecules per gelator, respectively immobile and total, assuming all the gelators contributing to the network. 

We eventually get the number of immobile molecules from the relation: 
\begin{equation}
n_{solvent,immo}=  \frac{\sigma_{solvent}\cdot n_{solvent} \cdot \left(\frac{I_{el,solid}}{I_{qens, solvent}}\right)_{Q=0} - \sigma_{\alpha manno}}{\sigma_{solvent}\cdot\left(\frac{I_{el, solid}}{I_{qens,solvent}}+1\right)_{Q=0}}
\end{equation}

\begin{figure}
\includegraphics[width=8cm]{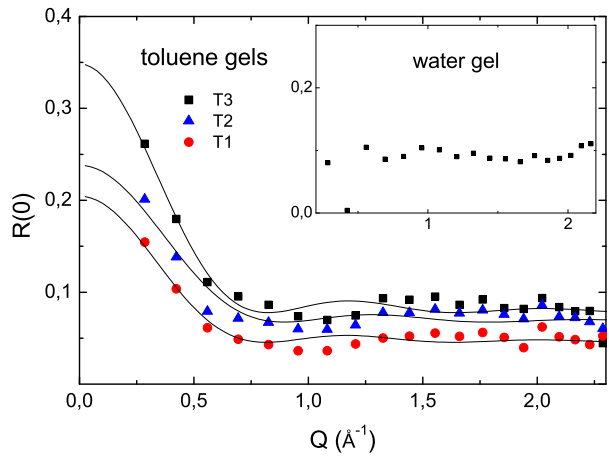}
\caption{Ratio between the elastic contribution and the total intensity measured (integrated over an energy range of [-5,1.8] meV) at $\lambda=5$ \AA$\,$  on IN5. The fits to the toluene data represent the rotational EISF (see text for details). The inset shows the same data for the water sample where no significant Q dependence is observed.} \label{ELASTIC_PIC}.
\end{figure}

These quantities are shown in Figure \ref{ELASTIC_PIC} for the four investigated samples. As can be observed for the toluene samples, a Q dependence is observed, showing that a relaxation is present even in the rigid network. Analyzing this Q dependence gives access to characteristic geometry and length scales of the undergoing dynamics in the solid. The variation of $R(0)$ as a function of $Q$ can be fitted with the structure factor given by an isotropic rotation: 
\begin{equation}
R(Q)=A \left( \frac{\sin r_s Q}{r_s Q} \right) ^2 +B
\label{eq:EISF}
\end{equation}
where A is the amplitude, B is the intensity corresponding to the atoms that do not participate to the relaxation, and $r_s$ is the radius of rotation. The equation \ref{eq:EISF} gives a reasonable description of the data and enables, beside extracting a radius of rotation, to extrapolate the data towards $Q=0$ and get the correct value for $R(0)$.

The number of solvent molecules $n_{solvent, immo}$ needed to reproduce $R(0)$ can then be extracted from the data. This number is the number of solvent molecules that are included in the rigid assembly. All the data are collected in Table \ref{tbl:tableIN5}, showing that an average of $6.2 \pm 2.5$ toluene molecules and $10 \pm3$ water molecules are immobilized on a timescale of 13 ps. Concerning the different toluene gels, the number of trapped solvent molecules does not, within the error bars, vary significantly with the gelator concentration. We will therefore assume the network constitution as concentration independent, and consider only the average number of immobilized toluene molecules. 
The simple isotropic rotation model, although providing a poor fitting of the oscillations at high Q due to the superposition of other relaxation processes, enables to extract a mean rotation radius. A value of $r_s=3.8\pm 0.2 $ \AA\, was extracted for the three samples, in good agreement with the rotation radius that could be expected from the size of the benzylidene moieties of the gelators or of the toluene molecules trapped in the network. The water sample, instead, do not exhibit any Q dependence of the elastic intensity, indicating that no dynamics of the solid part of the sample are detected on this timescale.

\begin{table}
   \begin{tabular}{| c |c |c |c |}
\hline
      sample &  solvent/gelator   & $R(0)$ & solvent immobilised\\
      &   num. mol. & norm. & num. mol.\\
	\hline     
      T1 &  52.1 & 0.20 $\pm$ 0.02 & 8.8 $\pm$ 3.8 \\
       T2 &  25.8  & 0.24 $\pm$ 0.02 & 4.5$\pm$ 0.8\\
       T3 &  18.5  & 0.35 $\pm$ 0.02 & 5.1$\pm$ 2.6\\
       H$_2$O - 1   & 180.2 & 0.1  $\pm$ 0.02 & 10 $\pm$ 3.0\\
\hline
   \end{tabular}
\caption{$\alpha$-manno gels used in the QENS experiments, ratio between elastic and total intensity $R(0)$ measured on IN5, and number of immobilised solvent molecule per gelator on 13 ps.}
\label{tbl:tableIN5} 
\end{table}

{\bf Sample fraction immobile up to 1.3 ns.} Quasi elastic measurements on the backscattering spectrometer IN16 (ILL) showed that no relaxation with characteristic time between 100 ps and 1.2 ns is present in the gel phase at 300 K. But the elastic intensity, recorded as a function of temperature in a so-called fixed window scan gives quantitative information on the solvent fraction that is effectively embedded in the rigid network \cite{Plazanet2006a,Plazanet2006}, in a similar way the elastic contribution did for the IN5 data. In such a scan, a calibration of the intensity can be performed by measuring at low enough temperature, typically 50 -250 K, where the whole sample is frozen. In this temperature range, the whole sample contributes to the elastic intensity, which decreases linearly with the temperature following a slope given by a pseudo-Debye-Waller factor, i.e. decrease of the intensity due to thermal agitation. The intensity is therefore proportional to the total number of molecules present in the sample, pondered by the suitable cross sections. The extrapolation of this slope at 300 K equals the intensity that would be measured if the whole sample was still solid, i.e. did not transit from the solid to the gel state (melted solvent). At 300 K, the bulk-like liquid fraction of the sample moves faster than the resolution and therefore do not contribute to the elastic peak any more. Like in the previous analysis, we assume that all the gelator molecules participate to the rigid network, for which we can calculate the elastic intensity. The difference of intensity between the experimental one and the one that would be given by the gelators can therefore be attributed to the solvent molecules that are immobilized together with the gelators. 

\begin{figure} 
\includegraphics[width=8cm]{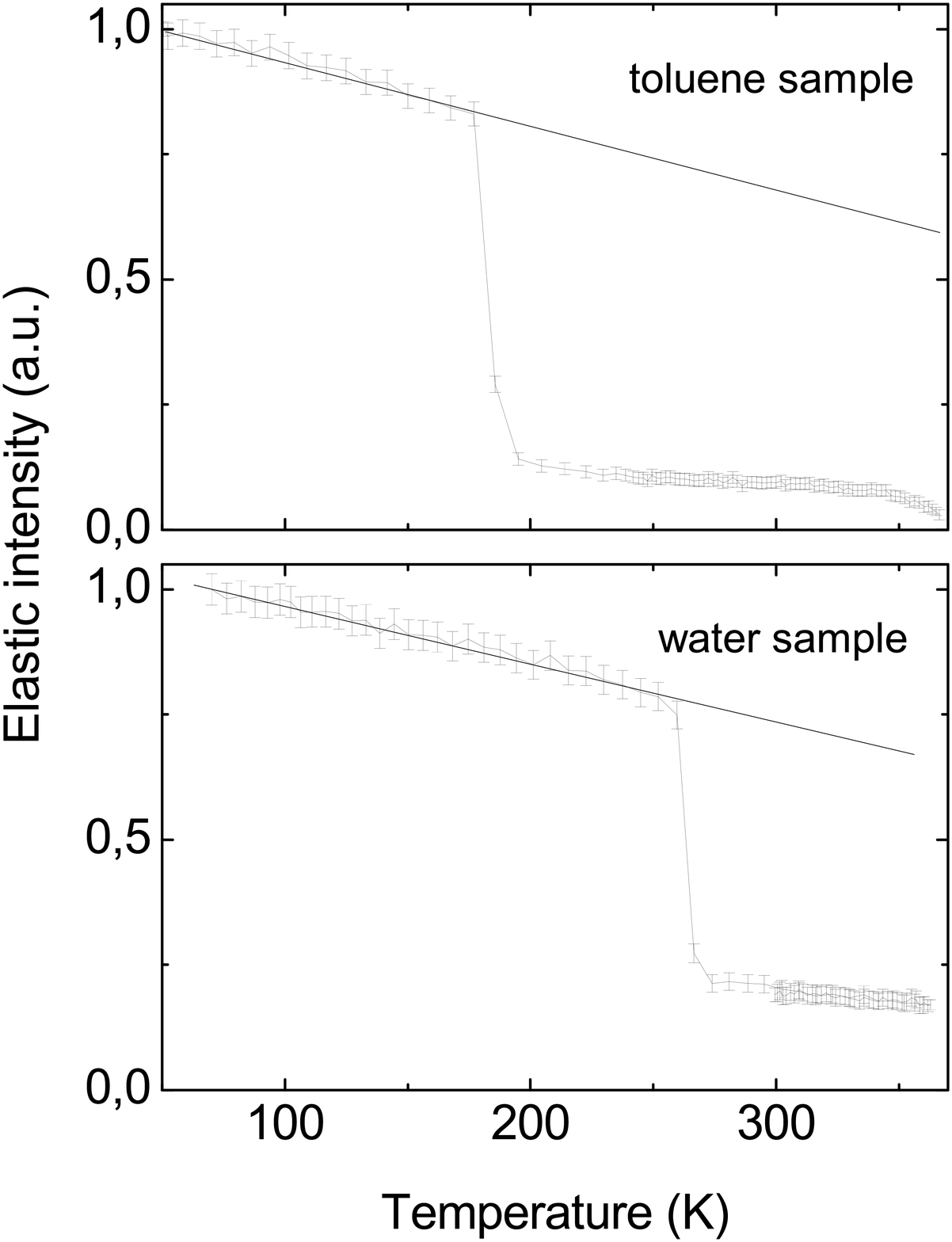}
\caption{Elastic scan as a function of temperature measured on IN16 (instrumental resolution 1 $\mu $eV) and integrated over the full Q range. The line indicates the extrapolation of the pseudo-Debye-Waller intensity toward high temperature, if the sample did not melt.} \label{figure_IN16}.
\end{figure}

We have $\displaystyle \frac{I_{DW}}{I_{exp}} =\frac{ [\sigma_{\alpha manno}+n_{solvent} \cdot \sigma_{solvent}]}{[\sigma_{\alpha manno}+n_{solvent, immo}\cdot \sigma_{solvent}]}$, where $I_{exp}$ is the measured intensity at 300 K and $I_{DW}$ is the intensity extrapolated at 300 K following the Debye-Waller slope. The data are represented in Figure \ref{figure_IN16} and the results are given in Table \ref{tbl:tableIN16}.

\begin{table}
   \begin{tabular}{| c |c |c |c |c|}
\hline
      sample &  num. solvent/gelator &  I$_{exp}$(300K)  & I$_{D.W.}$ & num. solvent immobilised\\
\hline
    T4  & 22.0 &1.25$\pm$0.25 & 8.6$\pm$0.25 & 1.23 $\pm$ 0.5\\
            H$_2$O - 2   & 75.2 & 2.60$\pm$0.25 & 9.60$\pm$0.25 & 14 $\pm$2\\
\hline
   \end{tabular}
\caption{$\alpha$-manno gels used in the IN16 experiments, $I_DW$, $I_exp$ and number of immobilized solvent molecule per gelator.}
\label{tbl:tableIN16} 
\end{table}

At 300K, this procedure gives $n_{solvent, immo}$ = 1.23$\pm$0.5 immobile toluene molecules in the $\alpha$-manno/toluene sample, and $n_{solvent, immo}$=14 $\pm$2  in the $\alpha$-manno/water gel. The result for water is slightly higher than the value obtained from the IN5 data corresponding to a shorter observation time (~13 ps for IN5 and ~1.3 ns for IN16), which obviously is unphysical. However taking into account the large error bars associated with the procedure followed to estimate the number of immobilized molecules, both results agree reasonable well and allow us to conclude that approximately 10-14 water molecules associate strongly to the gel network and exhibit therefore \textit{solid-like} dynamics instead of the bulk behavior found for the remaining water. In the case of toluene, the error bars are also large, but it seems clear that the number of immobilized molecules on a time scale of ~13 ps decreases when the observation period is increased by two orders of magnitude.

\section{Discussion}
The solvent dynamics measured by Quasi Elastic Neutron Scattering is relevant of the different nature of both gels. We indeed showed that only a few toluene molecules are trapped in the network formed by the $\alpha$-manno even on the picosecond timescale, while 10 to 14 water molecules are immobile up to a few nanoseconds in the hydrogel.
The different behaviour between water and toluene can be rationalized in the different nature of the interaction of both kinds of solvent with the gelator, already responsible for the different structural organization of the gels. In the case of water, H-bonds are formed between $H_2O$ and the polar groups of $\alpha$-manno. They are relatively strong and therefore fix water in a stable
manner, blocking their dynamics so no quasielastic signal from those water molecules is observed and they remain attached to the gel for times of the order of several ns or larger. For toluene, the interaction between the solvent and the gelator is likely through $\pi-\pi$ interactions between the aromatic rings of $\alpha$-manno and toluene. Due to steric effects this makes that the number of \textit{fixed} solvent molecules is smaller and that the dynamics of those molecules are not completely frozen, so that they retain some rotational freedom that produces the QENS signal assigned to the \textit{solid} and characterized by the EISF shown in Figure \ref{ELASTIC_PIC}. In this case, the weaker character of the interaction between solvent and gel makes that toluene can escape from the gel in a time scale of several hundreds of ps, so that the number of immobilized molecules is 5 on a time scale of about 13 ps and only ~1 molecule when the observation time is ~1.3 ns. In principle, we could expect that such escaping motion should appear as an additional relaxation slower than the bulk translational diffusion. However the relatively small number of molecules participating in such motion (~3-4 per gelator vs a total number of toluene molecules between 18 and 52 per gelator, depending on the concentration) together with the instrumental limitations (both in terms of resolution and energy range, as well as in terms of statistical accuracy), make this difficult to observe. 

It is interesting to compare the present hydrogel with sugar solutions, in which recent studies of hydration water dynamics show a slowing down of the dynamics with respect to pure water. The retardation factor varies depending on the study and the sugar concentration, and is estimated to about 3-6 in mono- and disaccharides \cite{Perticaroli2013, Lupi2012}, and does not exceed $\sim$ 15 for the first hydration layer \cite{Lupi2012a}. In our case, indeed, a slight slowing down of the water dynamics is observed by Time of Flight measurements, that could arise from the average of the bulk-like water and slowed down hydration water. However, the measured number of $\sim$10 to 15 molecules with dynamics slower than 1.2 ns indicates that the water molecules are more strongly involved in the rigid network than usual hydration water. 

The group of Tritt-Goc measured the diffusional dynamics of solvent in very similar gels made of methyl-4,6-O-(p-nitro-benzylidene)-$\alpha$-D-glucopyranoside as gelator, either in water or chlorobenzene \cite{Tritt-Goc2011,Kowalczuk2009}. This gelator is also an amphiphilic one, but dipole-dipole interactions are more important that in the $\alpha$-manno thanks to the nitro-substitute group. These forces act in favour of the gelation process, enabling gelation at lower concentration, and more stable gel phases than with the $\alpha$-manno. In water, the translational diffusion coefficients of gels with concentration ranging between 1-5 \% wt.  is similar to the one of bulk water, over distances of a few microns. They could not distinguish any component characteristic of water slower than the bulk \cite{Kowalczuk2009}. Similar studies, performed on a gel composed by the same gelator but in chlorobenzene, a slightly polar and aprotic solvent \cite{Tritt-Goc2011}, in the same concentration range, proved a significant slowing down of the solvent dynamics close to the gel surfaces. They were able to estimate to four the number of solvent molecules per gelator immobilized on the probed timescale, i.e. between 25 ns and 100 $\mu$s (10kHz and 40 MHz). Assuming both studies may be related, in sight of the slight diversity of gelators and solvent, we can conclude that exists an intermediate dynamical range on which the water molecules eventually start moving.

In conclusion, this study demonstrates that Quasi-Elastic Neutron Scattering can be used to characterize the solvent dynamics in a gel, on timescales complementary to NMR. The combination of both techniques indicate that several populations of solvent can be distinguished on different timescales: a few highly hindered solvent molecules with relaxation times in the microsecond range, and another population moving on the sub-nanosecond range.

\subsection*{Acknowledgements}
We thank the ILL for beam time allocation and are grateful to Jacques Ollivier for the help during the experiment on IN5.

\bibliography{paper-gel}

\newpage 
\begin{figure} 
\includegraphics[width=18cm]{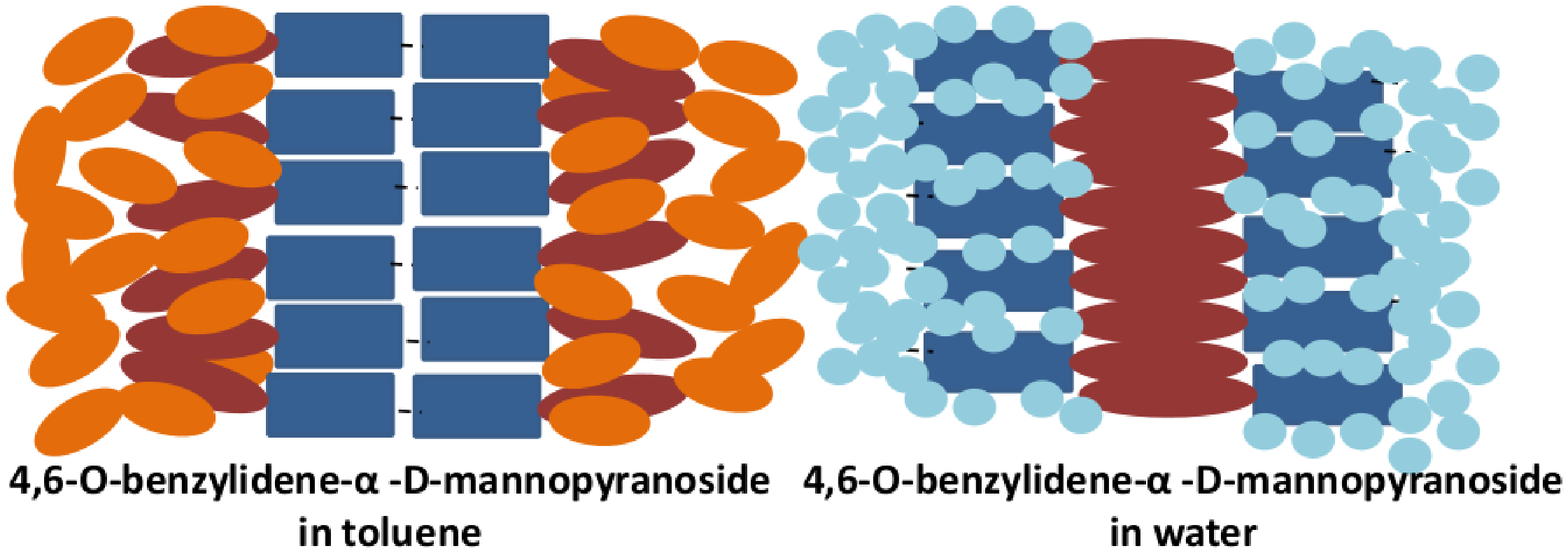}
\caption*{Graphical Table of Content.} \label{GTOC}.
\end{figure}

\end{document}